\documentclass[aps,prb,amsfonts,amssymb,twocolumn,amsmath,preprintnumbers,floatfix,showpacs]{revtex4}
\usepackage[dvips]{color}
\usepackage[dvips]{graphics}
\usepackage{graphicx}
\usepackage{amsbsy}
\usepackage{dcolumn}
\usepackage{bm}

\begin{document}

\title{Magnetic interference patterns in superconducting junctions:\\
Effects of anharmonic current-phase relations}

\author{Yu.\,S.~Barash}
\affiliation{Institute of Solid State Physics, Russian Academy of Sciences,
Chernogolovka, Moscow District, 142432 Russia}

\date{\empty}

\begin{abstract}
A microscopic theory of the magnetic-field modulation of
critical currents is developed for plane Josephson junctions with
anharmonic current-phase relations. The results obtained allow
examining temperature-dependent deviations of the modulation from the
conventional interference pattern. For tunneling through localized
states in symmetric short junctions with a pronounced anharmonic
behavior, the deviations are obtained and shown to depend on
distribution of channel transparencies. For constant transparency the
deviations vanish not only near $T_c$, but also at $T=0$. If Dorokhov
bimodal distribution for transparency eigenvalues holds, the averaged
deviation increases with decreasing temperature and takes its maximum
at $T=0$.
\end{abstract}


\maketitle

\section{Introduction}

Magnetic interference patterns in superconducting junctions originate
from quantum coherence of the superconducting state under the applied
magnetic field. They attract considerable experimental attention and
underlie effective studies of various problems of superconductivity
(see for example~\cite{Barone1982,vanHarlingen1995,TsueiKirtley2000,
Hilgenkamp2002,DellaRocca2005,Weides2006,Frolov2006,Chesca2008,
Strand2009,Kemmler2009}). Yet the corresponding microscopic theory
for short junctions is lacking and the present understanding of the
results is based mainly on the Ginzburg-Landau approach and the
tunneling limit. A microscopic extension of the results to the low
temperature region $T\ll T_c$ is required, since the Josephson
current is to a great extent controlled by discrete Andreev bound
states, which are not resolved within the Ginzburg-Landau theory.
Effects of finite transparencies of the transport channels are
intrinsically connected to the contributions of higher harmonics of
the supercurrent to the modulation, which, therefore, can be
adequately described only beyond the tunneling approximation. Due to
the absence of a corresponding microscopic theory, experimental data
on the interference patterns are analyzed in the literature partly
phenomenologically with reference to usual procedure firmly confirmed
for tunnel junctions near $T_c$.

When a transparency of plane junctions gets close to unity, there is
usually a crossover from the Josephson current to bulk
superconducting flow. Nonetheless, there are important plane
contacts, where the physics of weak links is still valid even in the
presence of highly transparent transport channels. This is the case
for long superconductor-normal metal-superconductor (SNS) fully
transparent junctions of various geometries, where interference
patterns have been studied in detail theoretically at arbitrary
temperatures and beyond the tunneling approximation~\cite{Antsygina1975,
Svidzinskii1982,Heida1998,Barzykin1999,Ledermann1999,Sheehy2003,
Cuevas2007,Bergeret2008,Mohammadkhani2008}. In particular,
the central Fraunhofer
peak in clean planar and long SNS junctions with fully transparent
interfaces has been found to get strongly distorted at low
temperatures. At $T=0$ it acquires a triangular
form~\cite{Antsygina1975, Svidzinskii1982}, which correlates with a
saw-toothed current-phase relation taking place under the same
conditions in the systems~\cite{Ishii1970, Svidzinskii1982}. This
example demonstrates that pronounced anharmonic current-phase
relations in superconducting junctions can entail significant
qualitative modifications in the corresponding magnetic interference
patterns.

Another characteristic weak link with a strongly anharmonic
current-phase relation is a short clean highly transparent point
contact, which in a fully transparent case reduces to the
Kulik-Omelyanchuk clean superconducting constriction~\cite{Kulik1977}.
Similar results also occur for tunneling through a
single localized state or for plane junctions, where resonant
electron tunneling takes place via individual localized states
homogeneously distributed over an insulating interface
(see~\cite{Kupriyanov1997,Kupriyanov2003, Golubov2004} and references
therein). In such systems an analytical description of the Josephson
current is possible at low densities of the transport channels with
arbitrary transparencies since the pair breaking effects are small
there.

In the present paper modulations of the critical current are
described based on a microscopic theory of Josephson junctions
generalized to the case of an applied magnetic field. An integration
of the modulated current over the plane of a rectangular junction is
carried out explicitly in a general form for arbitrary interface
transparencies. The answer is related to the phase dependent part of
the thermodynamic potential in the absence of the modulation, taken at
the field-dependent phase difference. The theory is applied to short
junctions with localized states homogeneously distributed over the
interface plane. A qualitative difference is demonstrated between
deviations of the modulated critical current from the Fraunhofer
pattern in junctions with constant transparency and Dorokhov bimodal
distribution of transparency eigenvalues. Junctions between isotropic
$s$-wave superconductors are considered below but the extension of
basic results to unconventional superconductors is straightforward.

\section{Modulation of the Josephson current}

\begin{figure}[!t]
\centering
\includegraphics[width=.75\columnwidth,clip=true]{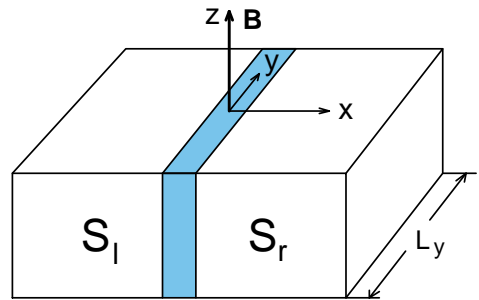}
\caption{Schematic diagram of the junction.}
\label{scheme}
\vspace{-0.2cm}
\end{figure}

Let superconducting electrodes ${\cal S}_l$ and ${\cal S}_r$ be thick
compared to the magnetic penetration depths $\lambda_{l(r)}$, while let
the thickness of the interlayer and the junction width be much less than
the coherence lengths $\xi_{l(r)}$ and the Josephson penetration
length, respectively. One takes the $x$ axis perpendicular to the
contact plane and the magnetic field applied along the $z$
axis:\,$\bm{B}(x)=B(x) \bm{e}_z$ (see fig.~\ref{scheme}).  It is
convenient to take the vector potential in the form $\mathbf{A}(x)=
A(x)\mathbf{e}_y$,\ $\mathrm{div}\,\mathbf{A}=0$, which coincides
with the gauge usually taken in describing the Meissner effect. In
contrast to the case of the Meissner effect, in Josephson junctions
the vector potential $\mathbf{A}(x)=A(x)\mathbf{e}_y$ does not vanish
everywhere in the depth of superconductors, where the screening
supercurrent $j_y(x)$ and the screened field $B(x)$ do vanish.
Indeed, a difference between asymptotic values of the vector
potential is associated with the magnetic flux $\Phi$ through the
junction:  $A_{+\infty}-A_{-\infty}= {\Phi}/{ L_y}$. Here $L_y$ is a
contact width along the $y$ axis.

Nonzero asymptotic values of the
vector potential can be excluded from microscopic equations by means
of the corresponding gauge-like transformation. Thus,
Bogoliubov amplitudes and order parameters
can be represented as $\tilde{u}^{r(\ell)}=u^{r(\ell)}
\exp\left[\frac{ie}{\hbar c}A_{\pm\infty}y\right]$,
$\tilde{v}^{r(\ell)}=v^{r(\ell)}
\exp\left[-\,\frac{ie}{\hbar c}A_{\pm\infty}y\right]$,
$\widetilde{\Delta}^{r(\ell)}=\Delta^{r(\ell)}
\exp\left[\frac{2ie}{\hbar c}yA_{\pm\infty}\right]$.
Here quantities $u^{r(\ell)}$, $v^{r(\ell)}$ and $\Delta^{r(\ell)}$
satisfy the equations where only the residual parts of the
vector potential $\widetilde{A}^{r(\ell)}(x)=A(x)-A_{\pm\infty}$ are
present in the right and the left superconducting regions,
respectively. The phases of the corresponding order parameters
$\widetilde{\Delta}^{r(\ell)}=\left|\Delta^{r(\ell)}\right|
\exp(i\widetilde\chi^{r(\ell)})$, $\Delta^{r(\ell)}=\left|\Delta^{r(
\ell)}\right|\exp(i\chi^{r(\ell)})$ are related as
$\widetilde{\chi}^{r(\ell)}(y)= \chi^{r(\ell)}+
\frac{2e}{\hbar c}A_{\pm\infty}y$. For nonzero
magnetic flux one gets $A_{+\infty}\ne A_{-\infty}$ and
the transformation does not reduce fully to fixing
a gauge since it differs in the two regions. Therefore,
after excluding constant asymptotic values of the vector potential
from microscopic equations, the quantities $A_{\pm\infty}$ enter
not only gauge-dependent phases of order parameters, of Bogoliubov
amplitudes and of Green's functions, but also a number of gauge-invariant
physical quantities. In particular, as the result of matching
corresponding solutions at the interface,
the phase difference $\widetilde\chi(y)=\widetilde\chi^{\ell}(y)-
\widetilde\chi^{r}(y)=\chi^{\ell}-\chi^{r}+
\frac{2|e|}{\hbar c}\left(A_{+\infty}-A_{-\infty}\right)y$ enters a
secular equation and influences the periodic phase-dependent spectrum of
interface Andreev states.

After performing the transformation in superconducting regions
the problem becomes formally more close to that of the
Meissner effect, since the residual parts of the vector potential
$\widetilde{A}^{r(\ell)}(x)$ vanish in the depth of the
superconductors together with $B(x)$ and $j_y(x)$. Similar to
the problem of the Meissner effect, $\widetilde{A}(x)$ in the given
gauge does not lead to any additional changes of phases of the order
parameters, even in a strongly nonlinear regime~\cite{ Agassi2006}.
For this reason the modulation of the critical current in the
microscopic theory is controlled by the spatial dependence of the
phase difference
\begin{equation}
\widetilde\chi(y)=\chi+2\pi(y/L_y)(\Phi/\Phi_0),
\label{tildechi}
\end{equation}
where $\chi=\chi^{\ell}-\chi^{r}$ and $\Phi_0=\pi \hbar c/|e|$ is the
superconducting flux quantum.

As the modulation period $L_y^B=\pi\ell_B^2/\overline{\lambda}=\Phi_0
/B(0)\overline{\lambda}$ is of a macroscopic scale, the
quasiclassical theory of superconductivity applies to a microscopic
study of the problem. Here $\ell_B=(\hbar c/|eB(0)|)^{1/2}$ is the
magnetic length and $\overline{\lambda}=d+\lambda_l+\lambda_r$,
where $d$ is the interlayer thickness. Within the quasiclassical
approximation, interface Andreev bound states are associated with
coupled incoming, reflected and transmitted trajectories, which cross
the interface at one and the same point. In the absence of the field,
Andreev bound states are degenerate with respect to the coordinates
$(y_0,z_0)$ of the reflection points, where parallel incoming
trajectories with given Fermi velocity $\mathbf{v}_f$ cross the
junction plane. The total supercurrent represents a sum of separate
contributions with various possible $\mathbf{v}_f$. When the external
magnetic field is present, the quasiclassical boundary conditions,
locally applied at each crossing point, result in lifting the
degeneracy due to $y_0$-dependence of the phase difference
$\widetilde\chi(y_0)$ across the interface. The periodic dependence
of the quasiparticle spectrum on the coordinate $y_0$ of the crossing
point, is the microscopic origin of the magnetic-field modulation of
the current. For describing the modulation, one should sum
(integrate) over $y_0$ the contributions to the current from
respective parallel trajectories for each given $\mathbf{v}_f$.

In the absence of the modulation, the phase-dependent part of the
thermodynamic potential of the junction can be represented as the
following sum over Matsubara frequencies
\begin{equation}
{\mathit\varOmega}_0(T,\chi)=-(T/2)\sum_{n=-\infty}^\infty\ln
D(i\varepsilon_n,\chi).
\label{varOmega0}
\end{equation}
The quantity $D(i \varepsilon_n,\chi)$ enters the secular equation
$D(\varepsilon,\chi) =0$ for eigenenergies of the system and can be
defined unambiguously~\cite{Beenakker1991,Brouwer1997}. In the
presence of spin degeneracy one gets
$D(\varepsilon,\chi)=D_\sigma^2(\varepsilon,\chi)$,
${\mathit\varOmega}=2{\mathit \varOmega}_\sigma$.

A variation of the thermodynamic potential with the phase difference for
a junction under the applied field $\delta{\mathit\varOmega}(T,\chi,
\Phi)$ is expressed via the variation $\delta{\mathit \varOmega}_0(T,
\chi)$ in the absence of the modulation:
\begin{equation}
\delta{\mathit\varOmega}(T,\chi,\Phi)=
\frac1{L_y}\int_{a-L_y/2}^{a+L_y/2}dy_0\delta{\mathit\varOmega}_0
\Bigl(T,\chi+\frac{2\pi\Phi}{\Phi_0}\frac{y_0}{L_y}\Bigr).
\label{deltaOmega}
\end{equation}
Here a rectangular plane junction is supposed to occupy the
space $(a-L_y/2,a+ L_y/2)$ along $y$ axis. The parameter $a$
determines a position of the interference pattern relative to the
junction edges.  Since the Josephson current and thermodynamic
potential satisfy the relation
\begin{equation}
I(T,\chi,\Phi)=\frac{-2e}{\hbar}\frac{d
}{d\chi}{\mathit\varOmega}(T,\chi,\Phi),
\label{I_varOmega}
\end{equation}
the integration of the current over $y_0$ can be explicitly carried out.
One obtains from eqs.~(\ref{deltaOmega}) and (\ref{I_varOmega})
\begin{equation}
I=\frac{e\Phi_0}{\pi\Phi\hbar}\left[{\mathit\varOmega}_0\Bigl(T,\chi_e-
\frac{\pi\Phi}{\Phi_0}\Bigr)-
{\mathit\varOmega}_0\Bigl(T,\chi_e+\frac{\pi\Phi}{\Phi_0}\Bigr)\right].
\label{I_varOmega0}
\end{equation}
Here $\chi_e=\chi+\frac{2\pi\Phi}{\Phi_0}\frac{a}{L_y}$
is the effective phase difference. As this follows from
eqs.~(\ref{varOmega0}) and (\ref{I_varOmega0}), the magnetic-field
modulation of the Josephson current at arbitrary temperatures and
transparencies is described by the expression
\begin{equation}
I(T,\chi_e,\Phi)=\frac{\displaystyle eT\Phi_0}{\displaystyle 2\pi\Phi\hbar}\sum_{n=-\infty}^\infty
\ln\left[\frac{\displaystyle D\left(i\varepsilon_n,\chi_e+\frac{\pi\Phi}{
\Phi_0}\right)}{\displaystyle D\left(i\varepsilon_n,\chi_e-\frac{\pi\Phi}{
\Phi_0}\right)}\right].
\label{IPhi1}
\end{equation}
Eq.~(\ref{IPhi1}) allows calculations of magnetic-field modulations
of critical currents, provided that the secular function
$D(i\varepsilon_n,\chi)$ is known for the junction in the absence of
the modulation. The secular function can take complex values and its
property $D(-i\varepsilon_n,\chi)=D^*(i\varepsilon_n,\chi)$ ensures
real values of thermodynamic potentials and the current.  Since
eqs.~(\ref{tildechi})-(\ref{I_varOmega}) underly the derivation of
Eq.~(\ref{IPhi1}) and have quite general character, eq.~(\ref{IPhi1})
applies to a variety of planar rectangular junctions with any
interfaces, including those between unconventional superconductors
and/or with magnetic interlayers.

In symmetric junctions the Josephson current is carried solely by
subgap states, for which
\begin{equation}
\delta{\mathit\varOmega}_0(T,\chi)=\delta\{-
T\sum_{i=1}^N\ln[2\cosh(E_{i}(\chi)\big/{2T})]\}.
\label{deltaOmega0}
\end{equation}
Here the sum is taken over Andreev state energies $E_{i}(\chi)>0$ of $N$
transport channels, which can depend on trajectory directions and
spin indices. According to eqs. (\ref{I_varOmega0}) and (\ref{deltaOmega0}),
\begin{equation}
I(T,\chi_e,\Phi)\!=\frac{\displaystyle eT\Phi_0}{\displaystyle \pi\Phi\hbar}
\!\sum\limits_{i=1}^N
\ln\left[\frac{\displaystyle \cosh\left({E_{i}\left(\chi_e+\frac{\pi\Phi}{
\Phi_0}\right)}\Big/{2T}\right)}{\displaystyle \cosh\left({E_{i}\left(\chi_e-
\frac{\pi\Phi}{\Phi_0}\right)}\Big/{2T}\right)}\right].
\label{IPhi2}
\end{equation}
Within its application domain eq.~(\ref{IPhi2}) agrees with
eq.~(\ref{IPhi1}). In particular, eq.~(\ref{IPhi1}) reduces to
eq.~(\ref{IPhi2}) in the simplest case,
when $D_{\sigma}(i\varepsilon_n,\chi)=\prod_{i=1}^{N}A_i
\left[\varepsilon_n^2+E_{i}^2(\chi)\right]$ and $A_i$ are
independent of $\chi$.

A phase difference $\chi_{e,c}(T,\Phi)$, which corresponds to the
modulated critical current $I_c(T,\Phi)=|I(T,\chi_{e,c}(T,\Phi),\Phi)
|$, satisfies the equation
\begin{equation}
I_0\Bigl(T,\chi_{e,c}(T,\Phi)+ \frac{\pi\Phi}{
\Phi_0}\Bigr)=I_0\Bigl(T,\chi_{e,c}(T,\Phi)-\frac{\pi\Phi}{\Phi_0}\Bigr),
\end{equation}
where $I_0(T,\chi)$ is the Josephson current in the absence of the
modulation. In the zero-field limit one obtains from eq.~(\ref{IPhi1})
or eq.~(\ref{IPhi2}) familiar general relations between the Josephson
current and the secular function or the spectrum of interface
Andreev bound states~\cite{Beenakker1991,Brouwer1997}. It follows from
eqs.~(\ref{IPhi1}) or (\ref{IPhi2}) that the current always vanishes under
the condition $D\left(i\varepsilon_n,\chi_e-\frac{\pi\Phi}{\Phi_0}
\right)=D\left(i\varepsilon_n,\chi_e+\frac{\pi\Phi}{\Phi_0}\right)$
or $E_{i}\left(\chi_e-\frac{\pi\Phi}{\Phi_0}\right)=E_{i}\left(\chi_e
+\frac{\pi\Phi}{ \Phi_0}\right)$. Hence, a $2\pi$-periodic
phase-dependent spectrum ensures positions of nodes of the modulated
Josephson current at $\Phi=n\Phi_0$,\enspace $n=\pm1,\pm2,\dots\,$,
irrespective of the phase difference. Since all even harmonics
also  vanish at $\Phi=\frac{2n+1}{2}\Phi_0$, for such values of the
magnetic flux the current is formed only by contributions from odd
harmonics. For small deviations $\delta
\Phi$ of the magnetic flux from $n\Phi_0 $, the current and, in
particular, its derivative with respect to the phase difference
always have opposite signs above and below each of the nodes.
This signifies that the positions of minima of thermodynamic
potentials as functions of the phase difference abruptly change by
$\pi$ at $\Phi=n\Phi_0$. Therefore, continuous $0$-$\pi$ transitions
of the interference origin take place with varying magnetic flux
through points $\Phi=n\Phi_0$ ($n=\pm1, \pm2,\dots\,$), where all
harmonics of the current vanish simultaneously. The $0$-$\pi$
transitions are known to take place, in particular, with varying
temperature or interface thickness in junctions with magnetic interlayers.
One can see that
such transitions also take place with varying magnetic field through a
junction. If the magnetic field, satisfying the relation
$n\Phi_0<\Phi<(n+1) \Phi_0$, is applied, then originally $0$ ($\pi$)
junctions either evolve to the $0$ ($\pi$) state with respect to
$\chi_e$ (for $n=0, \pm 2, \pm4\, \dots$), or turn into respective
$\pi$ ($0$) junctions (for $n=\pm1,\pm 3, \pm5, \,\dots$). This
concerns, in particular, the standard situation, when the Fraunhofer
pattern describes the modulation.

\section{Tunneling through localized states}

\begin{figure}[!t]
\centering
\includegraphics[width=.75\columnwidth,clip=true]{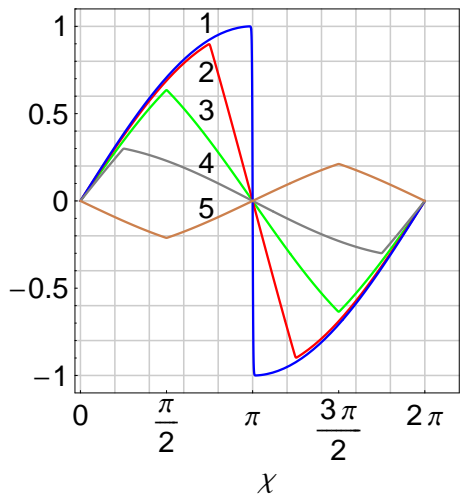}
\caption{The Josephson current $I(\chi,\Phi)/I_c(\Phi=0)$ as a
function of the phase difference in fully transparent (${\cal D}=1$)
symmetric junctions at $T=0$.  The current is normalized to the
critical current at zero-field and taken for various values of the
magnetic flux through the junction:  1.\,$\Phi=0$,\, 2.\,$\Phi=0.25
\Phi_0$,\, 3.\,$\Phi=0.5\Phi_0 $,\, 4.\,$\Phi=0.75\Phi_0$,\, 5.\,$
\Phi=1.5\Phi_0$.}
\label{cprh}
\vspace{-0.2cm}
\end{figure}

Consider further nonmagnetic short junctions between identical
$s$-wave superconductors, where tunneling via localized states with a
large broadening occurs. The influence of the screening current and
the magnetic orbital effects on the Josephson current is usually
negligibly small in such systems, so that the residual vector
potential can be disregarded. Then the spectrum of spin-degenerate
Andreev states takes the form $E_{i,\pm}(\chi)=\pm |\Delta| \sqrt{1-
{\cal D}_i\sin^2(\chi/2)}$, which formally coincides with the
spectrum of superconductor - insulator - superconductor point
contacts~\cite{Beenakker1991}. The transparency ${\cal D}_i$ is
described here by the Breit-Wigner resonance function, taken at the
energy of the $i$-th localized state~\cite{Kupriyanov1997,Kupriyanov2003,
Golubov2004,Glazman1989}.  The coefficient ${\cal D}_i$ can take any
value between $0$ and $1$, depending on the energy of the state and
its position $x_{i,0}$ across the interlayer. Near $T_c$ the order
parameter is small and, expanding all functions in eq.~(\ref{IPhi2})
in powers of $E_{+}/T_c$, one can keep there only the main quadratic
term. This leads to the relation $I_c(T,\Phi)=I_{cF}(T,\Phi)$, where
$I_{cF}(T,\Phi)$ describes the Fraunhofer pattern for the critical
current
\begin{equation}
I_{cF}(T,\Phi)=I_c(T,0)
\left|{\sin\left(\frac{\pi\Phi}{\Phi_0}\right)}\Big/{\left(
\frac{\pi\Phi}{\Phi_0}\right)}\right|.
\label{wkf1}
\end{equation}
In the particular case $I_c(T,0)|_{T\to T_c}=
\frac{|e||\Delta|^2}{4\hbar T_c}\sum_{i}{\cal D}_i$, where the sum is
taken over possible different $\mathbf{v}_f$.

At low temperatures arguments of hyperbolic functions in
eq.~(\ref{IPhi2}) are large. Using the respective asymptotic expressions
one obtains within a simplified model of constant $\cal D$:\,
$\cos\chi_{e,c}(0,\Phi)=\cos\chi_{c}(0,0)\cos\left(\frac{
\pi\Phi}{\Phi_0}\right)$. Here the zero-field phase difference is
$\cos\chi_{c}(0,0)=-(1-\sqrt{1-{\cal D}})^2/{\cal D}$. This solution
results in the zero-temperature critical current, which exactly
reduces to the Fraunhofer pattern eq.~(\ref{wkf1}) for any field value.
The zero-field critical current at $T=0$, which enters eq.~(\ref{wkf1})
as a factor and depends on ${\cal D}$, is found to take the form
$I_{c}(0,0)= (|e\Delta|/\hbar)(1-\sqrt{1-{\cal D}})$.

\begin{figure}[!t]
\centering
\includegraphics[width=.75\columnwidth,clip=true]{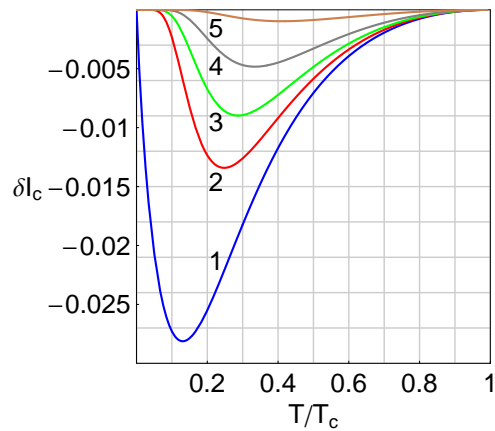}
\caption{Relative deviations $\delta I_c(T)$ as functions
of temperature in symmetric junctions, taken for
$\Phi=0.5\Phi_0$ and various transparencies: 1.\,${\cal D}=1$,\,
2.\,${\cal D}=0.95$,\, 3.\,${\cal D}=0.9$,\, 4.\,${\cal D}=0.8$,\,
5.\,${\cal D}=0.5$.}
\label{relsym}
\vspace{-0.2cm}
\end{figure}

Figure~\ref{cprh} shows current-phase dependences for various values
of the magnetic flux through symmetric fully transparent junctions in
question with a few identical transport channels at zero temperature.
The dependences involve significant contributions from a large number
of harmonics. Surprisingly, the conventional interference pattern for
the critical current in symmetric junctions takes place in this case.
Such behavior contrasts to what is known for long
superconductor-normal metal-superconductor junctions. Based on
eq.~(\ref{IPhi2}), one can calculate relative deviations $\delta
I_c(T,\Phi)=\Bigl(I_{c} (T,\Phi)-I_{cF}(T,\Phi)\Bigr)/|I_{c}(T,\Phi)
|$ of the critical current from the Fraunhofer values given by
eq.~(\ref{wkf1}). The quantity $\delta I_c(T,\Phi)$ vanishes
identically only in the tunneling approximation and/or near $T_c$.
Fig.~\ref{relsym} displays the deviation $\delta I_c(T,\Phi)$ as a
function of temperature, for $\Phi=0.5\Phi_0$ and various
transparency coefficients. At intermediate temperatures
eq.~(\ref{wkf1}) does not apply exactly, but the nonmonotonic
temperature dependent deviations due to higher harmonics are less
than few percent and vanish at $T=0$.

\begin{figure}[!t]
\centering
\centering
\includegraphics[width=.75\columnwidth,clip=true]{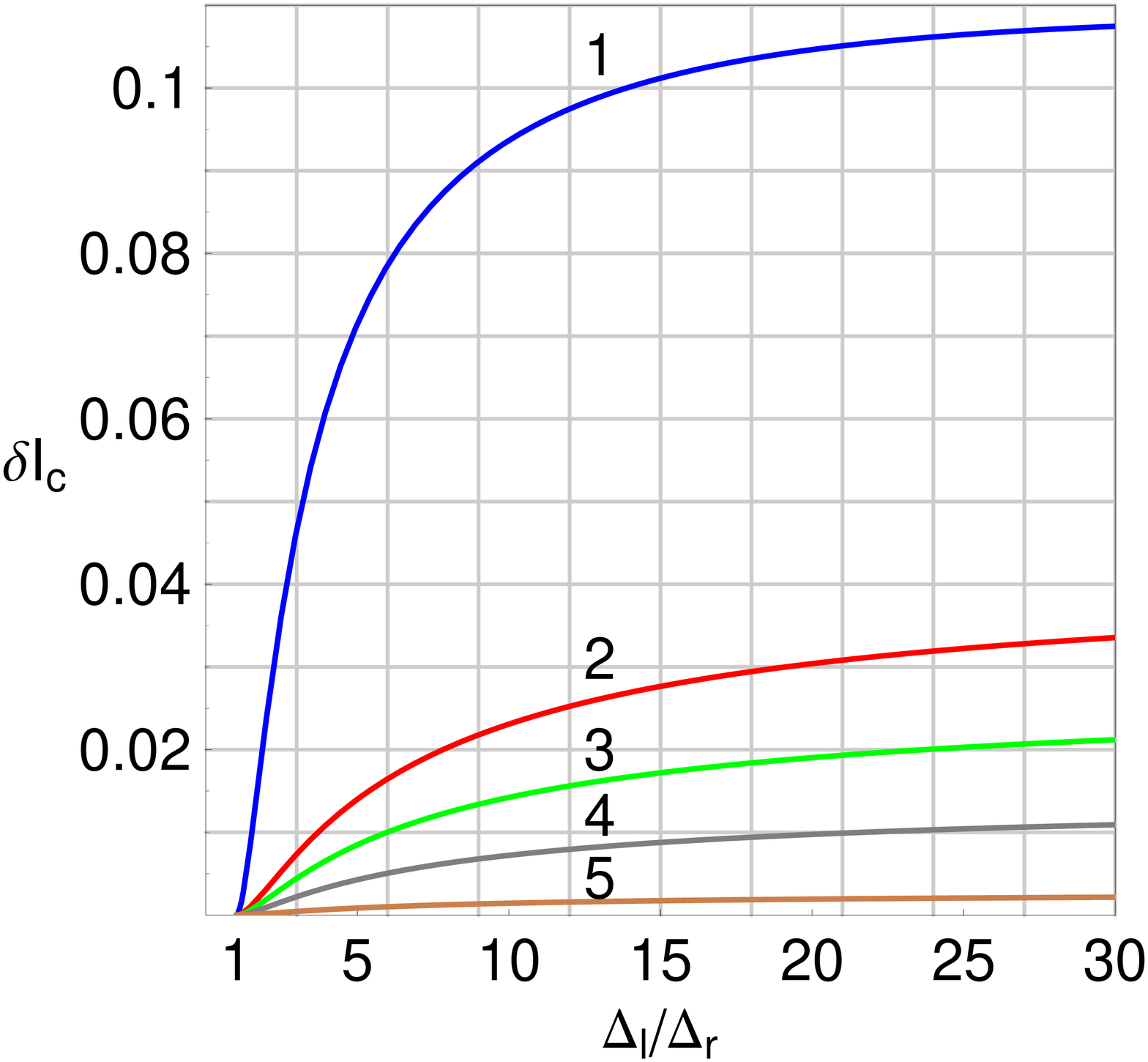}
\caption{Zero-temperature deviations
$\delta I_c(T=0)$ in asymmetric junctions as functions of
$\gamma=\left|\Delta_{l}/\Delta_r\right|$, taken for the same set
of parameters $\Phi$ and ${\cal D}$ as in fig.~\ref{relsym}.}
\label{relasym}
\vspace{-0.2cm}
\end{figure}

In asymmetric junctions zero-temperature deviations $\delta I_c$ do
not vanish, as this follows from eq.~(\ref{IPhi1}). They are shown in
fig.~\ref{relasym} as functions of the parameter $\gamma=\left|
\Delta_{l}/\Delta_r\right|$, which characterizes the junction
asymmetry. Since $\delta I_c$ does not vary with interchanging left
and right order parameters, one takes $\gamma\ge 1$. As can be seen,
$\delta I_c$ at $T=0$ is positive and not large, reaching about ten
percent at $\gamma=14$ and not exceeding eleven percent even at
$\gamma=30$.

\section{Multichannel effects}

In Josephson junctions with one transport channel or a few channels
with identical transparencies, the phase difference $\chi_{e,c}$,
which corresponds to the critical current, depends on the
transparency value due to anharmonic current behavior. In a
multichannel junction the critical current arises from an interplay
between channels with different transparencies and the resulting
value of $\chi_{e,c}$ can differ from those for the respective
one-channel junctions. For this reason the quantity $\chi_{e,c}$ in
strongly anharmonic junctions can noticeably influence the
dependence of the critical current on the transparency distribution
over transport channels.

\begin{figure}[!t]
\centering
\includegraphics[width=.75\columnwidth,clip=true]{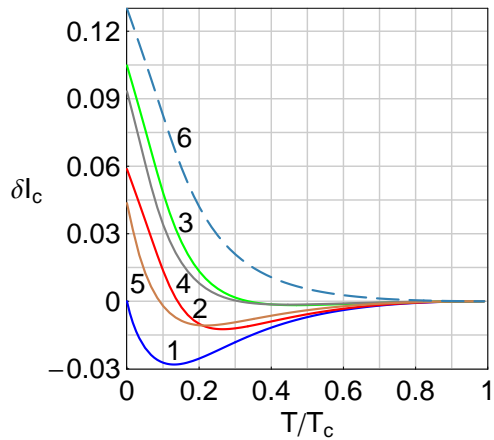}
\vspace{-0.5cm}
\caption{Relative deviations $\delta I_c$ as functions of
temperature, taken at $\Phi=0.5\Phi_0$ in symmetric junctions. Solid
curves correspond to junctions with two channels, where ${\cal D}_1=1$
and ${\cal D}_2$ takes the following values:\, 1.\,${\cal D}_2=1$,\,
2.\,${\cal D}_2=0.9$,\, 3.\,${\cal D}_2=0.5$,\, 4.\,${\cal D}_2=0.3$,\,
5.\,${\cal D}_2=0.1$. The dashed curve describes deviations in a junction
containing one channel with ${\cal D}_1=1$ and ten identical channels
with ${\cal D}_2=0.1$.}
\label{twochan}
\vspace{-0.2cm}
\end{figure}

A symmetric junction containing only two transport channels with
different transparencies ${\cal D}_{1}$, ${\cal D}_{2}$ represents
the simplest example for establishing multichannel effects. The
corresponding critical current can be found based on
eq.~(\ref{IPhi2}). Its relative deviations from eq.~(\ref{wkf1}) are
shown in fig.~\ref{twochan} as functions of temperature, for ${\cal
D}_{1}=1$ and various values of ${\cal D}_{2}$. Characteristic
qualitative features of the multichannel case follow from a
comparison of figs.~\ref{twochan} and~\ref{relsym}.
Figure~\ref{relsym} and the first curve in fig.~\ref{twochan} show
that at intermediate temperatures eq.~(\ref{wkf1}) slightly
overestimates the modulated critical current in symmetric junctions
with identical channels (constant transparency), but the deviations
vanish at zero temperature. Curves 2-5 in fig.~\ref{twochan}
demonstrate that an interplay between different channels can change
such situation. As a result, the critical current exceeds the
Fraunhofer value at low temperatures and, if ${\cal D}_2$ is not too
small, maximal deviations occur at zero temperature. A nonmonotonic
behavior of deviations at low temperatures taking place with the
parameter ${\cal D}_2$ arises from two competing reasons. If ${\cal
D}_2$ is not too small and decreases, then the variations of
$\chi_{e,c}$ from its value for constant transparency ${\cal D}_1=1$
increases and determines increasing deviations of the critical
current from the Fraunhofer one. On the other hand, for a channel
with sufficiently small and decreasing ${\cal D}_2$, its contribution
to the total critical current diminishes and the deviations are more
and more dominated by one channel with ${\cal D}_1=1$, i. e. they
decrease at zero temperature up to zero at ${\cal D}_2=0$.

The curve 6 in fig.~\ref{twochan} describes the critical-current
deviations in a junction containing ten identical channels with
${\cal D}_2=0.1$ and one fully transparent channel (${\cal
D}_{1}=1$). Contributions to the Josephson current from ten channels
with ${\cal D}_2=0.1$ and one channel with ${\cal D}_{1}=1$ are both
significant and jointly lead to more noticeable deviations of the
critical current.

\begin{figure}[!t]
\centering
\includegraphics[width=.75\columnwidth,clip=true]{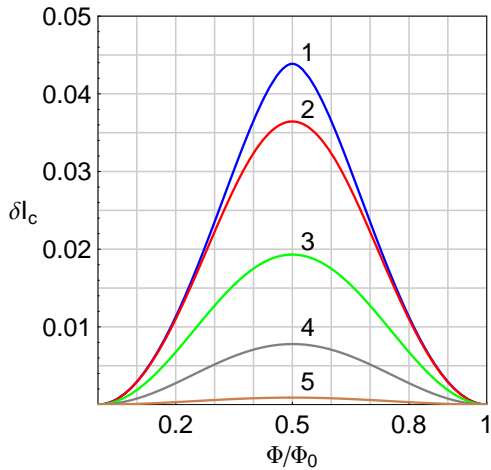}
\caption{Relative deviations $\delta I_c(T,\Phi)$ averaged over the Dorokhov
distribution of channel transparencies in symmetric junctions:
1.\,$T=0$,\, 2.\,$T=0.1 T_c$,\, 3.\,$T=0.2 T_c$,\, 4.\,$T=0.3 T_c$,\,
5.\,$T=0.5T_c$.}
\label{dorokhov}
\vspace{-0.2cm}
\end{figure}

For tunneling through broaden localized states the averaging of
$I_0(T,\chi)$ over bimodal Dorokhov distribution of transparency
eigenvalues leads to the current through dirty
constrictions~\cite{Golubov2004,Naveh2000}. The corresponding
thermodynamic potential takes the form
\begin{multline}
{\mathit\varOmega}_0(T,\chi)=\\ =\left({2\pi
\hbar^2T}\big/{e^2R_N}\right)\!\!\sum\limits_{\varepsilon_n>0}
\arcsin^2\left({\left|\Delta\right|\sin\frac{\chi}{2}}\big/{\sqrt{
\varepsilon^2_n+\left|\Delta\right|^2}}\right)
\end{multline}
and the modulated current $I(T,\chi,\Phi)$ is defined by
eq.~(\ref{I_varOmega0}). In this case the critical current exceeds
the Fraunhofer value, the deviation $\delta I_c(T,\Phi)$ increases
with decreasing temperature and takes its maximum at $T=0$, as is
seen in fig.~\ref{dorokhov}.

\section{Concluding remarks}

Experimental results for numerous short junctions are known to show,
as a rule, modulations of the standard type, if a spatial
distribution of the supercurrent density is not substantially
inhomogeneous~\cite{Barone1982}. Prominent exceptions include
combined $0$-$\pi$ junctions, vicinities of $0$-$\pi$ transitions and
special interface-to-crystal orientations of high-temperature or
other superconductors with anisotropic
pairings~\cite{vanHarlingen1995, TsueiKirtley2000,Hilgenkamp2002,
DellaRocca2005,Weides2006,Frolov2006,Goldobin2007,Chesca2008,
Strand2009,Kemmler2009}. The present calculations allow an extension,
in particular, to short junctions with interlayers possessing a
collinear magnetic order and/or between unconventional superconductors.
The developed approach can be also generalized to take account of the
current-induced magnetic field resulting in Josephson vortices in wide
junctions. These problems will be studied further and published elsewhere.

In conclusion, a microscopic theory of the magnetic-field modulation
of the critical current in Josephson junctions has been developed in
the present paper. As a generalization of basic microscopic results
in the absence of the magnetic field, the modulated Josephson current
is explicitly expressed via a secular function or, for symmetric
junctions, via a magnetic field dependent spectrum of Andreev
interface states. Temperature-dependent deviations of the modulated
critical current from the Fraunhofer pattern have been found for
short junctions with tunneling through localized electronic states.
The deviations depend on transparency distribution over transport
channels. For constant transparency the deviations vanish not only
near $T_c$, but also at $T=0$. Such behavior qualitatively differs
from what is known for long superconductor-normal
metal-superconductor junctions. Zero-temperature deviations are
found to take place in junctions between different superconductors
and in symmetric junctions containing channels with different
transparencies. If Dorokhov bimodal distribution of transparency
eigenvalues holds, the averaged deviation increases with decreasing
temperature and takes its maximum at $T=0$. It is shown that in a
number of junctions with a pronounced anharmonic current behavior,
the Fraunhofer pattern is only slightly distorted.

The support of RFBR grant 08-02-00842 is acknowledged.

\end{document}